\documentclass[a4paper]{jpconf}
\usepackage{graphicx}
\begin{document}
\title{Magnetotransport in graphene on silicon side of SiC}

\author{P.\ Va\v{s}ek$^1$, L.\ Smr\v{c}ka$^1$, P.\ Svoboda$^{1}$,
V.\ Jurka$^{1}$, M.\ Orlita$^{2,3}$, \\D.\ K.\ Maude$^{3}$,
W.\ Strupinski$^{4}$, R.\ Stepniewski$^{5}$, R.\ Yakimova$^{6}$}

\address{$^1$Institute of Physics ASCR, v.v.i., Cukrovarnick\'a 10, 162 53
  Praha 6, Czech Republic}
\address{$^2$Faculty of Mathematics and Physics,
  Charles University, Ke Karlovu 3, 121 16 Praha 2, Czech  Republic} 
\address{$^3$Laboratoire National des Champs Magn\'etiques
  Intenses, CNRS-UJF-UPS-INSA, 25, avenue des Martyrs, 38042 Grenoble,
  France}
\address{$^4$Institute of Electronic Materials Technology, Wolczynska 133, Warszawa, 01-919, Poland}
\address{$^5$Institute of Experimental Physics, Warsaw University, Hoza 69
  00-681 Warszawa, Poland}
\address{$^6$Link\"{o}ping University of Technology, S-581 83 Link\"{o}ping, Sweden}
\ead{vasek@fzu.cz}

\begin{abstract}
We have studied the transport properties of graphene grown on silicon
side of SiC.  Samples under study have been prepared by two different
growth methods in two different laboratories. 
Magnetoresistance and Hall resistance have been measured at
temperatures between 4 and 100 K in resistive magnet in  magnetic 
fields up to 22 T.  In spite
of differences in sample preparation, the field dependence of
resistances measured on both sets of samples exhibits two periods of
magneto-oscillations indicating two different parallel conducting
channels with different concentrations of carriers.  The
semi-quantitative agreement with the model calculation allows for 
conclusion that channels are formed by high-density  and
low-density Dirac carriers. The coexistence of two different groups
of carriers on the silicon side of SiC was not reported before.
\end{abstract}
\section{Introduction}
Epitaxial growth of graphene layers on the surface of
single-crystalline SiC, reported first in a seminal paper
\cite{Berg_2004}, is a well established way to the large scale
production of graphene for a future graphene based electronics. Many
modifications of this technology have been reported. In addition to
standard thermal decomposition of the upper layers of the SiC
substrate at temperatures above 1200$^\circ$C in various atmospheres,
chemical vapour deposition (CVD) on the SiC surface can also be
used. In the former case graphene arises by rearrangement of carbon
atoms left on the SiC surface after Si atoms sublimed out, in the
latter carbon atoms have to be added from outside by thermal
decomposition of suitable hydrocarbon gas on the surface. Growth
details can be expected to influence transport properties of graphene
even on SiC substrates of the same crystallographic orientation, due
to various types of active defects and/or character of the interface
between the graphene and the substrate. In the present paper we report on the
experimental study of the  and Hall resistances on two
series of samples, prepared from wafers epitaxially grown on Si-side
of SiC, using either Si sublimation or CVD methods.
\section{Experiments}
\subsection{Graphene growth}
Two series of samples have been employed in this study. One of them,
denoted as G141, was produced at the University of Link\"oping,
Sweden, on the semi-insulating 6H-SiC(0001) wafer by the
Si-sublimation method in an inductively heated furnace. The growth was
carried out under highly isothermal conditions at 2000$^{\circ}$C and
at an ambient argon pressure of 1 bar \cite{Viro_2010}.

Samples in the second series stem from a wafer 699, supplied by ITME
Warsaw, Poland. The graphene growth was performed in a commercially
available horizontal CVD hot-wall reactor (Epigress V508), inductively
heated by an RF generator. Semi-insulating 4H-SiC (0001) substrates
were used and propane gas served as the precursor. The substrates were
etched in hydrogen and propane mixtures prior to carbon
deposition. Precautions had been taken to prevent sublimation of Si
atoms in conditions of high temperature ($\approx $1600$^{\circ}$C)
and a low Ar pressure \cite{Strup_2011}.

Raman spectroscopy was used to confirm the presence of graphene.
\subsection{Lithography}
Samples were patterned by standard photolithographic techniques. Hall
bars with a 100 micrometers wide conducting channel, two current
contacts and six contacts for measurement of the longitudinal
magnetoresistance and the Hall voltages were formed by oxygen plasma
etching. The separation of potential leads was 300 micrometers. 
The aspect ratio of the sample (width/length)  was either 1/3 or 1/6, 
depending on the pair of contacts employed. 
 The macroscopic size indicates that the
sample area covers a large number of terraces on the SiC surface and
inhomogeneity in the graphene layer cannot be excluded. Ohmic TiAu
contacts were prepared by the electron beam assisted deposition and
lift-off technique. 
\begin{figure}[b]
\begin{minipage}[t]{0.47\linewidth}
\hspace{-6mm}
\includegraphics*[width=1.15\linewidth]{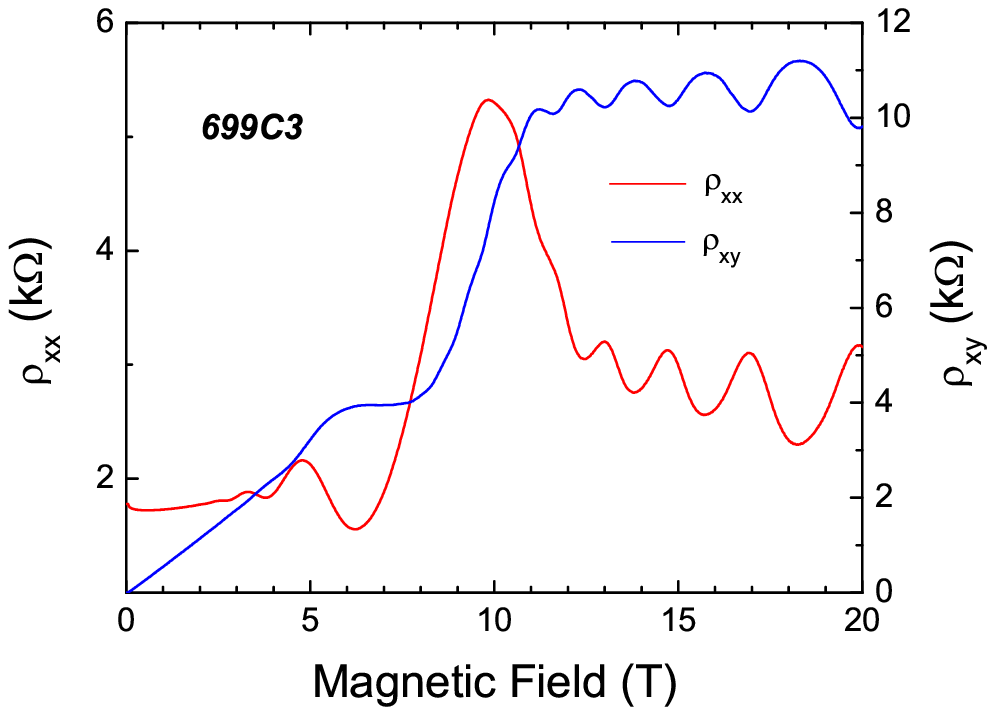}
\caption{\label{Fig.1} Longitudinal (red curve) and Hall (blue curve)
magnetoresistances mesured on a sample prepared in ITME Warsaw,
Poland.
$N_1=9.5\times 10^{11}$ cm$^{-2}$, $N_2=1.1\times 10^{13}$
cm$^{-2}$.}
\end{minipage}%
\hspace{0.06\linewidth}
\begin{minipage}[t]{0.47\linewidth}
\hspace{-8mm}
\includegraphics*[width=1.15\linewidth]{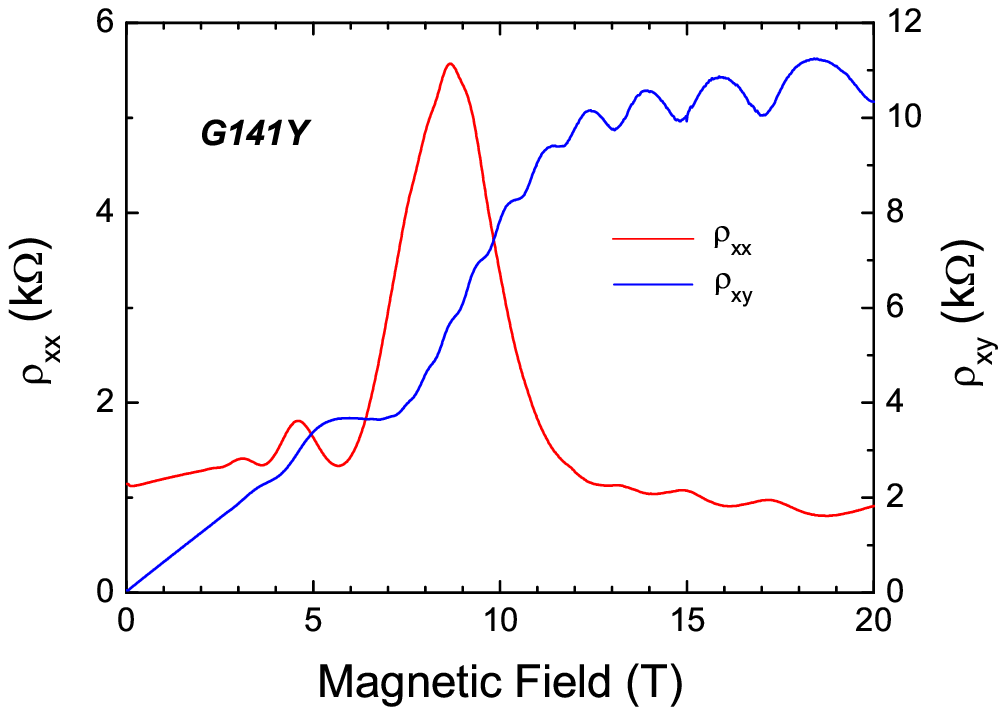}
\caption{\label{Fig.2}Longitudinal (red curve) and Hall (blue curve)
magnetoresistances mesured on a sample prepared in University of
Link\"oping, Sweden.
 $N_1=8.9\times 10^{11}$ cm$^{-2}$, $N_2=1.4\times
10^{13}$ cm$^{-2}$.}
\end{minipage}
\end{figure} 
\begin{figure}[hbp]
\begin{minipage}[t]{0.47\linewidth}
\hspace{-6mm}
\includegraphics[width=1.15\linewidth]{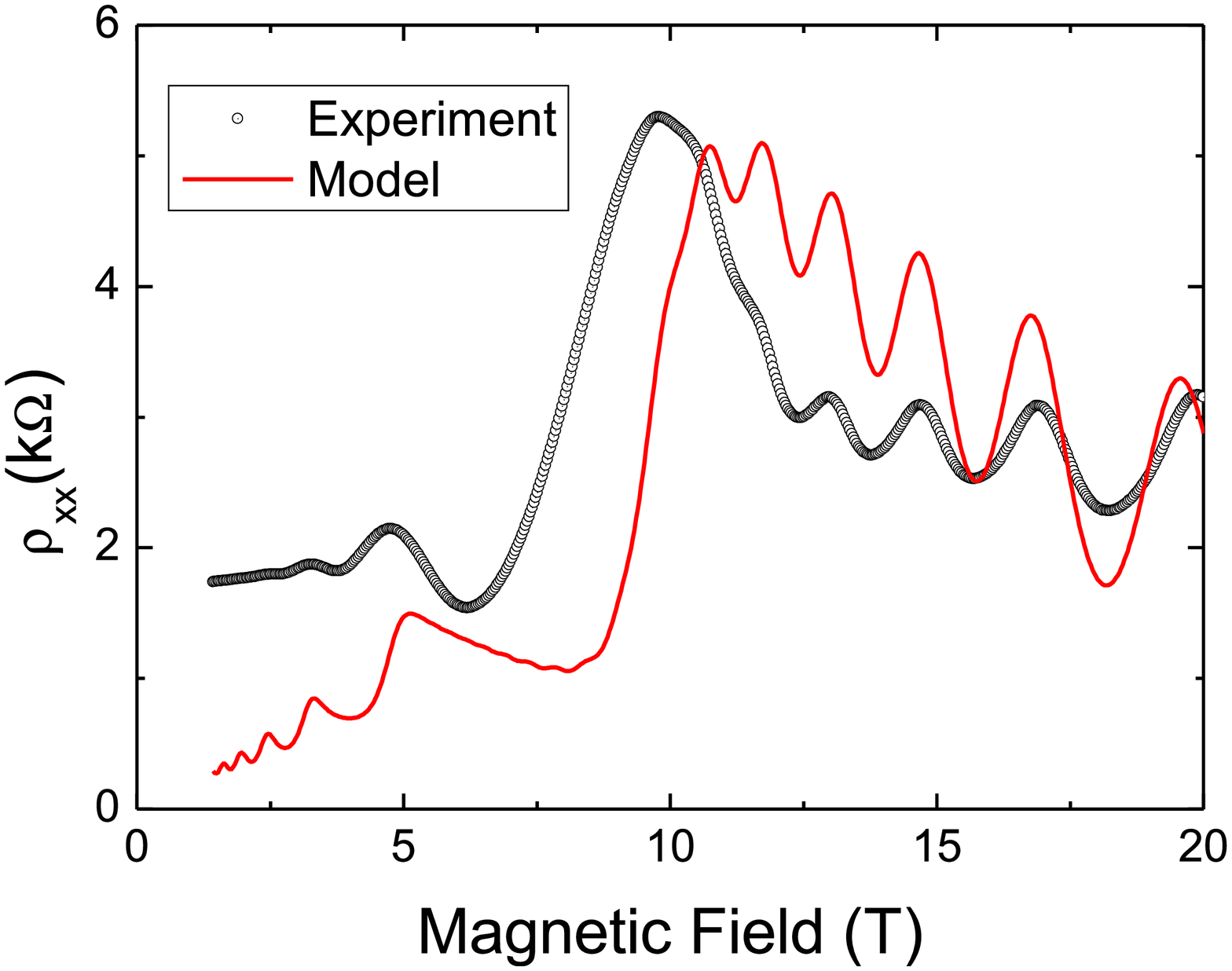}
\caption{\label{Fig.3} Experimental and model curves  of 
longitudinal magnetoresistance of a sample 699.}
\end{minipage}%
\hspace{0.06\linewidth}%
\begin{minipage}[t]{0.47\linewidth}
\hspace{-6mm}
\includegraphics[width=1.15\linewidth]{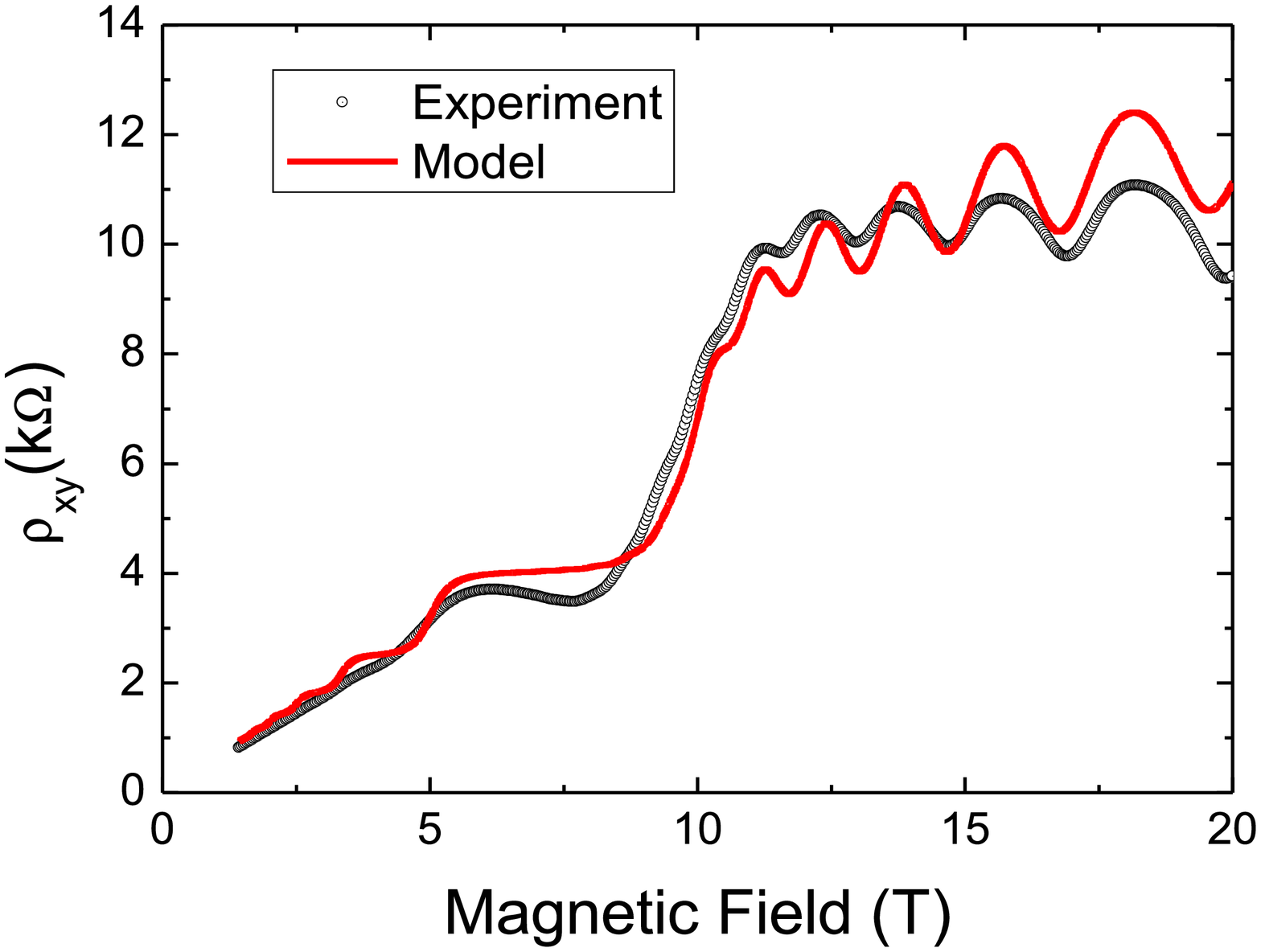}
\caption{\label{Fig.4}Experimental and model curves of Hall
  magnetoresistance of a sample 699.}
\end{minipage}
\end{figure}
\subsection{Magnetotransport measurement} 
Magnetoresistance $\rho_{xx}(B)$ and Hall resistance $\rho_{xy}(B)$
have been simultaneously measured by a 4-probe low frequency ($f$ =
10.66 Hz) AC lock-in method. Excitation current up to 100 nA was taken
from the output of the lock-in amplifier using 10 M$\Omega$ or 100
M$\Omega$ load resistors. Most experiments were performed with samples
directly immersed in a bath of liquid helium at 4.2 K. For the
investigation of the temperature dependence up to 100 K a VTI has been
used with a calibrated Cernox thermometer as a temperature
sensor. Transverse magnetic fields up to 22 T were generated by a 10
MW Bitter magnet. Admixtures of $\rho_{xx}$ to $\rho_{xy}$ due to
contacts misalignment was removed by symmetrization of all traces
measured in positive and negative magnetic fields.

All  experiments were realized in LNCMI CNRS Grenoble, France.
\section{Results and discussion}
The dominant feature observed on the magnetoresistance (MR) curves
$\rho_{xx}(B)$ shown in Fig.~\ref{Fig.1} and Fig.~\ref{Fig.2} is the
presence of two types of oscillations, both periodic in $1/B$ but with
different periods. Simultaneously, the Hall resistance $\rho_{xy}(B)$
oscillates at fields above 10 T. All these oscillations disappeared
after the samples were rotated into parallel configuration with
respect to applied magnetic field, which confirms that the
oscillations originate in two-dimensional sheets of charge carriers.

Such a behavior cannot stem from a single conducting sheet and
therefore two independent channels have to exist. From the periods of
oscillations in longitudinal magnetoresistance $\rho_{xx}(B)$ we can
estimate charge carriers concentrations in both channels. Denoting the
channels responsible for the ,,low field'' and ,,high field''
oscillations by indices 1 and 2, respectively, we obtained for the
sample 699C3 (Fig.~\ref{Fig.1}) the values $N_1 =
9.5\times10^{11}$cm$^{-2}$ and $N_2 = 1.1\times10^{13}$ cm$^{-2}$.
The same analysis of the data presented in Fig.~\ref{Fig.2} gives the
concentrations $N_1 = 8.9\times10^{11}$cm$^{-2}$ and $N_2 =
1.4\times10^{13}$cm$^{-2}$.

Apart from the modulation by the oscillations at higher fields, the
Hall curves for both samples clearly exhibit QHE plateaux at
$\rho_{xy}(B)= (h/\nu e^2)$. Absence of the plateaux at $\nu$ = 4
confirms, that the ,,low field'' oscillations can be attributed to the
single layer graphene. The identification of the second conducting
layer with substantially higher carrier concentrations $N_2$ seems to
be less straightforward, the Hall plateaux cannot be reached in
magnetic fields at our disposal.  Assuming the same character of both channels,
we could compare experimental data for the sample 699C3 in
Fig.~\ref{Fig.1} with calculations based on the model
described in \cite{Burgt_1995}, including the appropriate description of the
conductivity tensor for graphene  \cite{Zheng_2002}, and assuming two parallel
conductivity channels in the sample.

In these calculations, concentrations
$N_1$ and $N_2$, together with the half-widths $\Gamma_1= 2.9$ meV and
$\Gamma_2= 6$ meV of Landau levels serve as fitting parameters.
The results of the best fit are presented in Figs.~\ref{Fig.3} and
\ref{Fig.4}. The model describes the data reasonably well, in
particular those in Fig.~\ref{Fig.4} for the Hall
resistance. Moreover, the best fit provides concentrations $N_1 =
1.2\times10^{12}$cm$^{-2}$ and $N_2 = 1.1\times10^{13}$cm$^{-2}$. These
agree quite well with the concentrations derived directly from the
periods of both types of MR oscillations observed on the same sample.

\begin{figure}[t]
\begin{minipage}[t]{0.47\linewidth}
\hspace{-6mm}
\includegraphics*[width=1.15\linewidth]{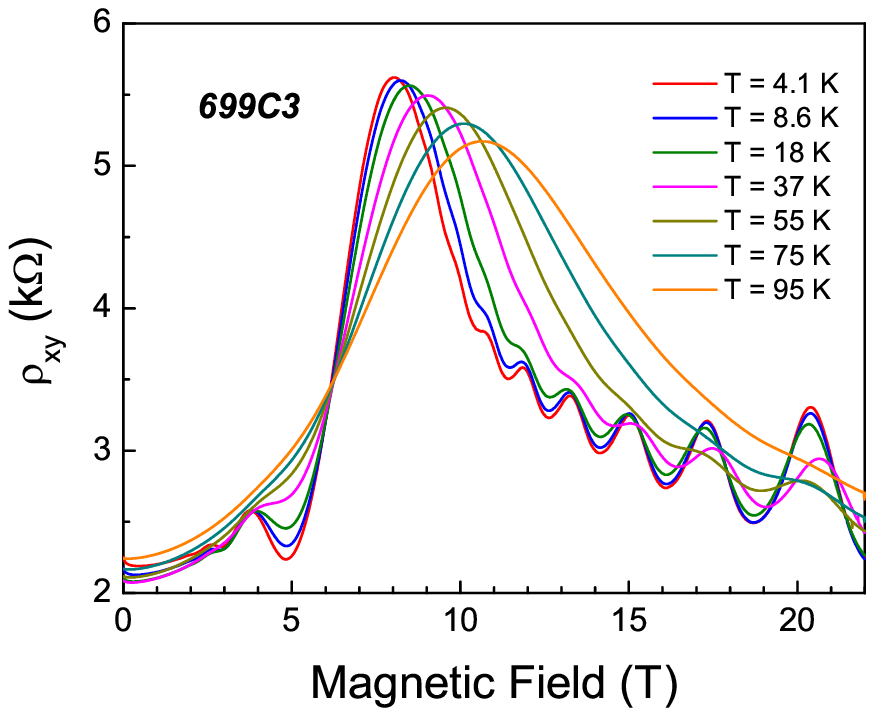}
\caption{\label{Fig.5}Longitudinal magnetoresistances mesured at
various temperatures on sample prepared in ITME Warsaw, Poland.}
\end{minipage}%
\hspace{0.06\linewidth}
\begin{minipage}[t]{0.47\linewidth}
\hspace{-8mm}
\includegraphics*[width=1.15\linewidth]{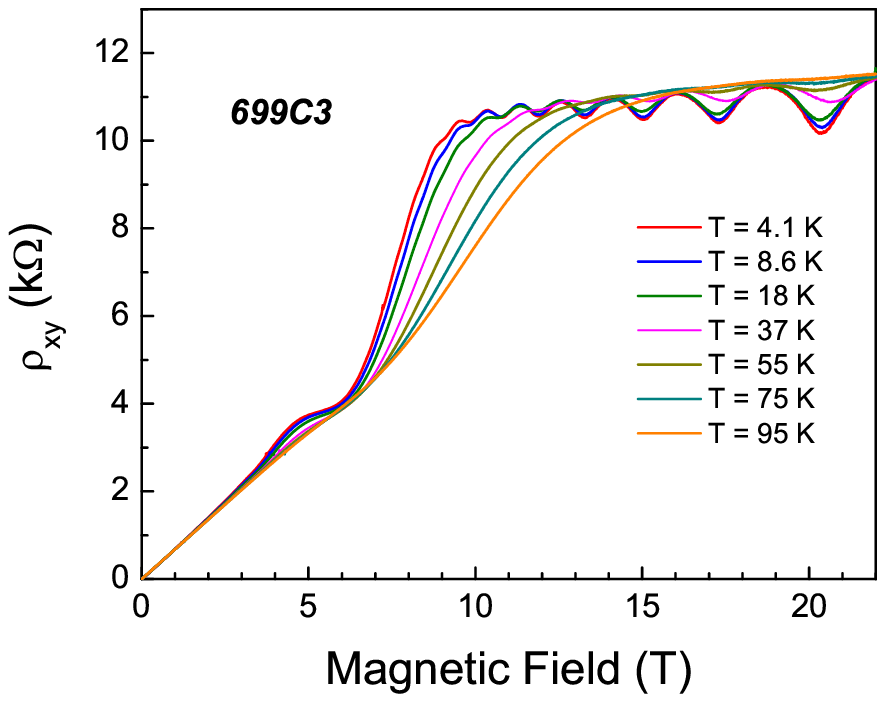}
\caption{\label{Fig.6} Hall magnetoresistances mesured at
various temperatures on sample prepared in ITME Warsaw, Poland.}
\end{minipage}
\end{figure}

In order to further characterize the two conducting channels, we have
measured temperature dependence of $\rho_{xx}(B)$ and $\rho_{xy}(B)$
up to about 100 K. The results can be seen in Fig.~\ref{Fig.3} and
Fig.~\ref{Fig.4}.  Both types of oscillations persist up to quite high
temperatures, which is characteristic property of graphene.
Temperature damping of the amplitude of oscillations provides an
estimate of the cyclotron effective mass $m_c^*$ of the charge
carriers \cite{Tan_2011}. It implies a reliable subtraction of the
monotonous ,,background'' part, which is not easy here, in particular
for the ,,low field'' oscillations. Nevertheless, in highest fields,
where the resistivity $\rho_{xx}(B)$ from the channel responsible for
the ,,low field'' SdH oscillations approaches zero, we got values of
$m_c^*$ that agree reasonably well with those derived from the
concentrations $N_i$. The latter stem from the expression $m_c^*=
\hbar/v_F$ for the effective mass of the single-layer graphene
\cite{Tan_2011}. Taking $v_F = 1.1\times 10^6$ m/s  \cite{Tan_2011}
and the concentrations $N_1$ and $N_2$ given in caption to
Fig.~\ref{Fig.1}, we get $m_{c1}^{*}= 0.018\, m_0$ and $m_{c2}^{*}= 0.061\,
m_0$, respectively.  The curves presented in Figs.~\ref{Fig.5} and
\ref{Fig.6} were obtained more than a year later than those shown in
Fig.~\ref{Fig.1} (see \cite{Orli_2011}). This confirms the stability of the
sample properties in time.

We can conclude, that our results for the two wafers produced by
widely different technologies can be interpreted as a superposition of
oscillations from two parallel two-dimensional conducting channels
with substantially different carrier concentrations and apparently
also mobilities.

It seems to be surprising, since different type of substrates together
with different growth technology can be expected to provide quite
different morphology of the epitaxial graphene. Detailed study of the
graphene formed on SiC under various environments \cite{Shriva_2012} has shown no
apparent differences between results for 6H-SiC and 4H-SiC
wafers. Substantial differences can however arise between samples
produced by the purely sublimation method and the CVD assisted one. In
the former case the existence of an interfacial (buffer) layer is well
established and it influences the electronic transport properties of
graphene [9], most probably through the concentration and mobility of
the charge carriers. But due to the bonding of the buffer layer to the
Si atoms in the substrate it loses its graphene-like linear band
structure near the Fermi level \cite{Hibi_2012} and cannot thus generate SdH
oscillations characteristic for the single layer graphene. In the CVD
assisted method, the structure of the interface between graphene and
the SiC substrate is less understood, but it probably also influences
only the concentration and mobility of the charge carriers in
uppermost graphene layers.  It is important that both technologies
produce graphene, that is far from being flat and homogeneous on the
scale larger than a few microns. Graphene seems to nucleate on the
edges of the terraces and grow further in islands until they fuse into
a layer, which is supported by the diffusion of the carbon atoms along
the surface \cite{Hibi_2012}. It can be expected, that not only the morphology and
thickness \cite{Hibi_2012}, but also the transport properties can be different for
graphene at the terraces and their edges \cite{Grod_2012}.  Therefore we
speculate, that the surface area of our samples is broken into small
interconnected regions with either low or high concentration of
carriers, which form a network of parallel and/or series resistors. It
may be a combination of the monolayer graphene at the terraces and
bilayer graphene at their edges \cite{Shriva_2012}, but the electron transport
measurement itself can hardly reveal the detailed microstructure of
the epitaxial graphene on SiC substrate. Explanation of the origin of
dual SdH oscillations observed in our samples thus requires further
investigation.

\ack{The support of the European Science Foundation EPIGRAT project
(GRA/10/E006, 670 and 671/N-ESF-EPI/2010/0), GACR No.~P204/10/1020 
and No.~P108/11/0894 projects,
programme ,,Transnational
access'' contract No.~228043-EuroMagNET II-Integrated Activities,
AVCR research program AVOZ10100521, the Academy of Sciences of the
Czech Republic project KAN400100652  are acknowledged.}


\section*{References}
\begin{thebibliography}{9}
\bibitem{Berg_2004}
Berger C,   Song Z,  Li T,  Li X,  Ogbazghi  A Y,  Feng R,
 Dai Z,  Marchenkov A N,  Conrad E H,  First P N and
 de Heer  W A   2004 {\it J. Phys. Chem.} B {\bf 108} 19912
\bibitem{Viro_2010}
Virojanadara C, Zakharov A A, Yakimova R and Johansson L I
2010 {\it Surf. Sci.}  604, L4
\bibitem{Strup_2011}
Strupinski W, Grodecki K, Wysmolek A, Stepniewski R, Szkopek T,
Gasskell P E, Gr\"uneis A, Haberer D, Bozek R, Krupka J and Baranowski J
M 2011 {\it Nano Lett.} {\bf 11} 1786 
\bibitem{Burgt_1995}
van der Burgt M, Karavolas V C, Peeters F M, Singleton J, Nicholas R J, 
Herlach  F, Harris J J, Van Hove  M and Borghs G 1995 {\it Phys. Rev.} B {\bf 52} 12218 
\bibitem{Zheng_2002}
Zheng Y and  Ando T 2002  {\it Phys. Rev.} B {\bf 65} 245420
\bibitem{Orli_2011}
Orlita M, Maude D K, Goncharuk N A,  Jurka V,  Va\v{s}ek P,
 Svoboda P,  Smr\v{c}ka L,  Strupinski W, Yakimova R and  Jawad-ul-Hassan 
2011 {\it Annual Report LNCMI Grenoble-Toulouse} eds F Duc and D Maude
CARBON ALLOTROPES p.10
\bibitem{Tan_2011}
Zhenbing Tan, Gangling Tan, Li Ma,  Liu G T, Lu L and  Yang C L 2011
{\it Phys. Rev.} B {\bf 84}  115429
\bibitem{Shriva_2012}
Shrivastava N, Guowei He, Luxmi, Mende P C, Feenstra R M and Yugang
Sun 2012 {\it J. Phys. D: Applied Phys.} {\bf 45} 154001
\bibitem{Hibi_2012} 
Hibino H, Tanabe S, Mizuno S and Kageshima H 2012 {\it J. Phys. D:
Applied Phys.} {\bf 45} 154008
\bibitem{Grod_2012}
Grodecki K,   Bozek R,  Strupinski W,  Wysmolek A,  
Stepniewski R and Baranowski J M 2012 {\it Appl. Phys. Lett.} {\bf100} 261604 
 

\end{thebibliography}
\end{document}